# Scanning nano-SQUID with single electron spin sensitivity

Denis Vasyukov[1*†], Yonathan Anahory[1†], Lior Embon[1], Dorri Halbertal[1], Jo Cuppens[1], Lior Ne'eman[1], Amit Finkler[1], Yehonathan Segev[1], Yuri Myasoedov[1], Michael L. Rappaport[1], Martin E. Huber[1,2] and Eli Zeldov[1*]

**One of the critical milestones in the intensive pursuit of quantitative nanoscale magnetic imaging tools is achieving the level of sensitivity required for detecting the field generated by the spin magnetic moment $\mu_B$ of a single electron[1,2]. Superconducting quantum interference devices (SQUIDs), which are traditionally the most sensitive magnetometers, could not hitherto reach this goal because of their relatively large effective size (of the order of 1 µm) and planar geometry[3-6]. Here we report self-aligned fabrication of nano-SQUIDs with diameters as small as 46 nm and with an extremely low flux noise of 50 n$\Phi_0$/Hz$^{1/2}$, representing two orders of magnitude improvement in spin sensitivity, down to 0.38 $\mu_B$/Hz$^{1/2}$. In addition, the devices operate over an unprecedentedly wide range of magnetic fields with 0.6 $\mu_B$/Hz$^{1/2}$ sensitivity even at 1 T. We demonstrate magnetic imaging of vortices in type II superconductor that are 120 nm apart and scanning measurements of AC magnetic fields down to 50 nT. The unique geometry of these nano-SQUIDs that reside on the apex of a sharp tip allows approaching the sample to within a few nm, which paves the way to a new class of single-spin resolved scanning probe microscopy.**

    A DC SQUID consists of a superconducting loop with two Josephson junctions or weak links. Its operation is based on the fact that as a result of quantum interference the maximum dissipationless current $I_c$ that can flow through the SQUID is periodic in the magnetic flux $\Phi$ through the loop[7], with a period of the flux quantum $\Phi_0=h/2e$ ($h$ is the Planck's constant and $e$ is the electron charge). The flux sensitivity of a SQUID is determined by its intrinsic flux noise with a typical value of the order of $S_\Phi^{1/2} = 1$ µ$\Phi_0$/Hz$^{1/2}$. In order to achieve a sensitive magnetometer with a low magnetic field noise, $S_B^{1/2} = S_\Phi^{1/2} / A$, SQUIDs are commonly designed to have pickup loops with large effective area $A$. In recent years there has been growing interest in the development of nano-SQUIDs for the study of quantum magnetism and for nanomagnetic imaging[3-12]. In this case, field sensitivity is compromised for the benefit of spatial resolution and sensitivity to magnetic dipoles. Since the magnetic field of a dipole has a $1/r^3$ dependence on the distance $r$ from the dipole source, by reducing the loop size and allowing closer proximity to the source, nano-SQUIDs become extremely sensitive detectors of small magnetic moments, with one of the ultimate goals being sensitivity to a single electron spin or 1 $\mu_B$.

    Most micro- and nano-SQUIDs are fabricated using planar lithographic techniques, which permit the integration of pickup and excitation loops[4]. Our devices, in contrast, are fabricated on the apex of a hollow quartz tubes pulled into a very sharp pipette, the ideal geometry for scanning probe microscopy (Fig. 1). Superconducting films are deposited in three self-aligned steps resulting in two superconducting leads connected to a superconducting loop (see Fig. 1a and Methods). The SQUID geometry consists of a loop with two weak links in the form of Dayem bridges in the gap regions between the leads.

Similar devices[12] made of Al have the drawback of working only at temperatures below 1 K. Also the large coherence length $\xi$ and penetration depth $\lambda$ of Al prevent realization of sensitive devices with characteristic dimensions below 0.1 µm. Here we report on new ultra-small nano-SQUIDs-on-tip (SOTs) made of Nb and Pb, which have significantly shorter $\xi$ and $\lambda$ and as a result can operate at much higher magnetic fields and temperatures, yield more than an order of magnitude smaller effective area, and achieve more than two orders of magnitude better spin sensitivity.

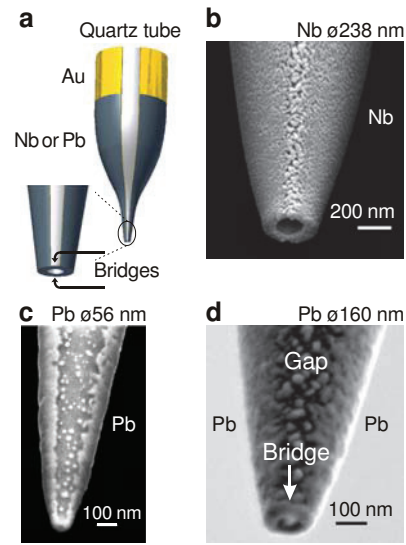

**Figure 1 | Scanning electron microscope images of SQUID-on-tip devices. a,** Schematic illustration of the pulled quartz tube with two Nb or Pb superconducting leads connected to Au electrodes. Inset: magnified view showing the superconducting loop on the apex of the tip. The bridges that reside in the gap regions between the leads form the two weak links of the SQUID. **b, c, d,** Images of the Nb SOT with an effective diameter of 238 nm, Pb SOT of ø56 nm, and Pb SOT of ø160 nm respectively. The islands that are visible in the gap region between the leads do not percolate and thus do not short the two leads.

A SEM image of a Nb SOT is shown in Fig. 1b. The Nb was deposited by e-gun evaporation in order to have a point-like source since magnetron sputtering is incompatible with the SOT fabrication geometry; a dedicated UHV system was developed with a high-power e-gun and an in-situ rotatable stage. A 10 nm thick buffer layer of AlO$_x$ was added onto the quartz tips to prevent contamination from the overheated glass during the Nb deposition (see Methods). In order to overcome the very high room-temperature surface mobility of Pb, which results in island growth and consequently percolation only at large thicknesses, we have developed a high-vacuum system with an in-situ rotatable $^4$He cryostat for thermal deposition of Pb onto cryogenically cooled tips. We have also optimized the pipette pulling process to reach apex diameters as low as 40 nm. Figures 1c and 1d show SEM images of two Pb SOTs with effective

[1]Weizmann Institute of Science, Department of Condensed Matter Physics, Rehovot 76100, Israel.
[2]Department of Physics, University of Colorado, Denver, CO 80217, USA.
*email: denis.vasyukov@gmail.com, eli.zeldov@weizmann.ac.il.
[†]These authors contributed equally to this work.

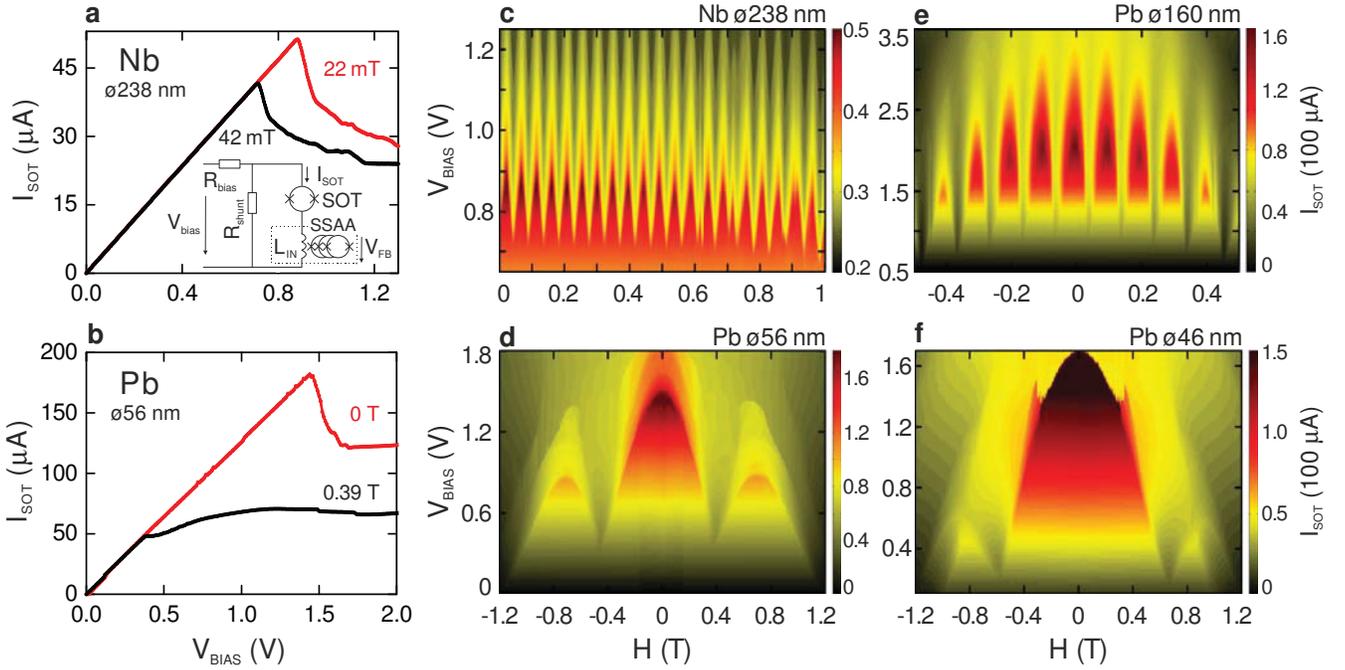

**Figure 2 | Quantum interference patterns and I-V characteristics of several SOT devices at 4.2 K. a,** Nonhysteretic I-V characteristics of ø238 nm Nb SOT at two fields corresponding to the maximum (red) and minimum (black) $I_c$. The maximum of $I_c$ is shifted away from zero magnetic field as a result of an asymmetry in the two junctions of the SOT. Inset shows the simplified measurement circuit (see Methods). **b,** I-V characteristics of the ø56 nm Pb SOT at fields of maximum and minimum $I_c$. **c,** Current through the Nb SOT (color code) vs. input voltage $V_{bias}$ and applied field $H$ showing quantum interference oscillations with a period corresponding to an effective SOT diameter of 238 nm. **d,** Quantum interference patterns of a Pb SOT with an effective diameter of 56 nm. Only three oscillation periods are obtained before $I_c$ is fully suppressed at $\pm H_{c2}$ = 1.2 T. **e,** Current oscillations in the ø160 nm Pb SOT. **f,** Current oscillations in the ø46 nm Pb SOT. Since the oscillation period is 1.27 T only the central period is clearly visible, while the neighboring lobes are truncated by vanishing of the critical current at $\pm H_{c2}$ = 1.2 T. The oscillation periods of the SOTs are in excellent agreement with the diameters measured by SEM (Fig. 1).

diameters of 160 and 56 nm, respectively.

Figure 2 compares the quantum interference patterns and current-voltage characteristics (I-Vs) of Nb ($T_c \cong 8.7$ K) and Pb ($T_c \cong 7$ K) SOTs measured using the circuit shown in Fig. 2a (all measurements reported in this paper are obtained at 4.2 K). The Nb SOT shows pronounced critical current $I_c$ oscillations up to exceptionally high field[13,14] of over 1 T (Fig. 2c) with a period corresponding to an effective loop diameter of 238 nm. The interference patterns of Pb SOTs are characterized by very large modulations of $I_c$ with an envelope that vanishes at 1.2 T. SOT operation at such high fields is possible because all the characteristic sizes of our devices are of the order of $\xi$ and the thin leads are aligned almost parallel to the applied field. Figures 2d to 2f show three Pb SOTs with effective diameters of 160, 56 and 46 nm, respectively. The smallest SOT has an extremely large oscillation period of 1.27 T (Fig. 2f) and thus only the central lobe is visible in full in the field range of $-H_{c2}$ to $+H_{c2}$ (where $H_{c2}$ = 1.2 T is the upper critical field). This device has an effective area that is some 20 times smaller than any previously reported[4,10,12] nano-SQUID. The individual $I_c$ oscillations of the Pb SOTs of all sizes display the usual sinusoidal field dependence of $I_c$ characteristic of Josephson junctions and short bridges[7]. In contrast, the Nb SOT oscillations show saw-tooth behavior indicating a linear current-phase relation, characteristic of Dayem bridges that are long[15] relative to $\xi$. We thus conclude that $\xi$ in the Nb apex ring is significantly shorter that in Pb, consistent with the lower critical current and higher apparent $H_{c2}$ in the Nb devices. Typical I-Vs are presented in Figs. 2a and 2b showing very large $I_c$ modulation in the Pb SOTs. The I-Vs display pronounced negative differential resistance, yet they are nonhysteretic due to the use of an effective voltage bias circuit in Fig. 2a (see Methods and Supplementary Fig. S1).

The flux noise spectra of the devices are shown in Fig. 3a. The white noise of the Nb SOT is $S_\Phi^{1/2} = 3.6\ \mu\Phi_0/\text{Hz}^{1/2}$, comparable[7] to conventional SQUIDs. In contrast, the Pb SOTs display flux noise as low as $S_\Phi^{1/2} = 50\ \text{n}\Phi_0/\text{Hz}^{1/2}$. This is one of the lowest flux noise levels ever achieved[16-18] in SQUIDs. The ultimate flux noise in SQUIDs is determined by the quantum noise[8,19] $S_{\Phi Q}^{1/2} = (\hbar L)^{1/2}$ where $L$ is the loop inductance. In our devices $L$ is governed by the kinetic inductance, e.g. $L_k \approx 5.8$ pH in the ø56 nm Pb SOT derived from $I_c$ modulation depth[19] leading to $S_{\Phi Q}^{1/2} = 12\ \text{n}\Phi_0/\text{Hz}^{1/2}$ (see Supplementary Information). The measured flux noise is thus 4.2 times higher than the quantum noise limit. Since $L_k$ is some two orders of magnitude higher than the geometric inductance, further significant improvement in the flux noise can be anticipated by reducing the kinetic inductance through optimization of material parameters. The corresponding field noise is also very low for nano-SQUIDs, reaching $S_B^{1/2} = 5.1\ \text{nT/Hz}^{1/2}$ in the ø160 nm Pb SOT. The smaller Pb SOTs shows very low *1/f* noise of the order of 0.5 $\mu\Phi_0/\text{Hz}^{1/2}$ at 1 Hz, see Fig. 3a. An outstanding feature of the SOTs, however, is their extremely low spin noise. For a spin in the center and perpendicular to the SQUID loop, the spin noise in units of $\mu_B$/Hz$^{1/2}$ is given by $S_n^{1/2} = S_\Phi^{1/2} \, r / r_e$ where $r$ is the loop radius and $r_e = 2.82 \times 10^{-15}$ m is the classical electron radius[4]. The Nb SOT achieves spin noise of $S_n^{1/2} = 154$ $\mu_B$/Hz$^{1/2}$, comparable to the best results yet obtained[4,5,12,20]. Figure 3b shows the spin noise spectra in the smaller Pb SOTs at various fields. Remarkably, the ø46 nm SOT reaches sensitivity of 0.38 $\mu_B$/Hz$^{1/2}$ that represents more than two orders of magnitude improvement[4,5,12] over any previously reported SQUIDs. Moreover, in sharp contrast to conventional SQUIDs and a number of recent optical readout magnetometers that can operate only at low fields[21-25], the SOTs can operate and remain very sensitive also at high fields. The ø56 and the ø46 nm Pb SOTs display spin noise of 0.75 and 0.6 $\mu_B$/Hz$^{1/2}$, respectively, at an unprecedented field of 1 T.

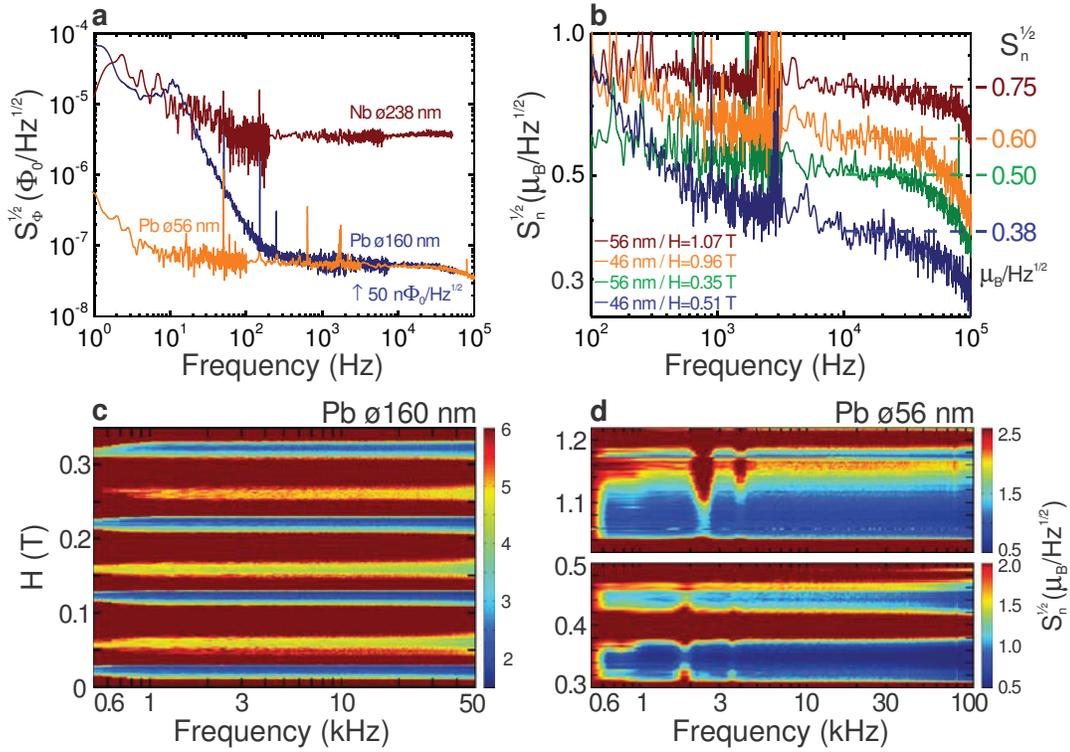

**Figure 3 | Flux and spin noise spectra of the SOTs at 4.2K. a,** Flux noise of the ø238 nm Nb SOT (red) and of ø160 nm (blue) and ø56 nm (orange) Pb SOTs. The white noise level of the Pb SOTs reaches 50 n$\Phi_0$/Hz$^{1/2}$. **b,** Spin noise of ø56 and ø46 nm Pb SOTs calculated for a dipole residing at the center and perpendicular to the SOT loop. The ø46 nm SOT displays spin sensitivity of 0.38 $\mu_B$/Hz$^{1/2}$ at 0.5 T and 0.6 $\mu_B$/Hz$^{1/2}$ at 1 T. The roll off of the spectra above 30 kHz is due low pass filtering. **c,** Spin noise spectra (color code) vs. frequency and applied field for the ø160 nm Pb SOT at a constant $V_{bias}$=1.7 V. Noise below 5 $\mu_B$/Hz$^{1/2}$ is colored in blue to yellow; field regions with noise above 6 $\mu_B$/Hz$^{1/2}$ are colored in brown. **d,** Spin noise spectra (color code) vs. frequency and $H$ for Pb ø56 nm SOT at a constant $V_{bias}$ = 0.55 V (bottom) and 0.3 V (top). Noise values below 1.5 $\mu_B$/Hz$^{1/2}$ are colored in blue to green in the field range of 0.3 to 0.5 T (bottom) and below 2 $\mu_B$/Hz$^{1/2}$ in blue to yellow in the range of 1.02 to 1.22 T (top).

Despite the fact that the SOT design does not include a feedback coil, as commonly used[7] in planar SQUIDs, the SOT can be operated over a wide range of fields because of its periodic response and tunability of bias parameters. The color code in Fig. 3c shows the noise spectra in the ø160 nm Pb SOT demonstrating a noise level of less than 5 $\mu_B$/Hz$^{1/2}$ over wide ranges of fields up to 0.35 T. Similarly, the ø56 nm Pb SOT displays noise below 1 $\mu_B$/Hz$^{1/2}$ over a range of fields between 0.3 and 0.5 T and below 2 $\mu_B$/Hz$^{1/2}$ between 1 and 1.2 T. Additional field ranges, including sensitive operation even at zero field, can be obtained by properly designing the SOT diameter, bias conditions, and asymmetry as shown in the Supplementary Information.

Since the SOT can approach and scan a sample within a few nm of the surface[12,26], an even higher spin sensitivity and enhanced spatial resolution can be obtained near the loop edges rather than in the center[8,27]. Figures 4a and 4b show the calculated flux coupling to the ø160 nm SOT from a single electron spin at a safe distance of 10 nm. If the spin is located near the loop edges, the flux in the loop is 65 and 88 n$\Phi_0$ for spins oriented perpendicular and parallel to the loop plane, respectively. With the Pb SOT noise of 50 n$\Phi_0$/Hz$^{1/2}$, this coupling corresponds to spin sensitivities of 0.77 and 0.57 $\mu_B$/Hz$^{1/2}$ for the two orientations, as compared to 1.4 $\mu_B$/Hz$^{1/2}$ for a perpendicular spin in the center of the loop. There is, in addition, an excellent spatial resolution of 20 nm as shown in Figs. 4d and 4e. Similarly, Figs. 4c and 4f show the flux in the loop owing to a single perpendicular spin 10 nm below the ø46 nm SOT giving rise to 0.53 $\mu_B$/Hz$^{1/2}$ sensitivity.

In order to demonstrate the spatial resolution of the SOTs, we have imaged vortices in a Nb thin film, as shown in Figs. 5a and 5b. The distance between vortices, $a = (2\Phi_0 / \sqrt{3}B)^{1/2}$ (for the hexagonal lattice), and the local modulation of the magnetic field decrease rapidly with increasing field $B$. In addition, the amplitude of the field modulation decays as $\exp(-z/a)$ where $z$ is the height above the surface of the superconductor. As a result, all previous scanning magneto-sensitive imaging of vortices was limited to very low fields[8,28-31] of up to a few mT. Figure 5b shows scanning SOT imaging of vortices at applied fields of up to 0.2 T. The distance between the two nearest vortices in Fig. 5b is only 120 nm.

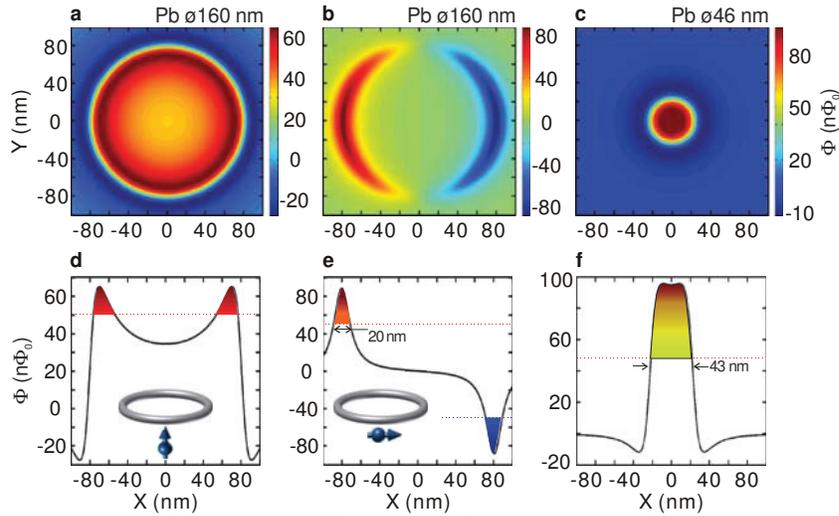

**Figure 4 | Calculated magnetic flux in the SOT loops vs. XY position of a single spin at 10 nm below the SQUID plane. a,** 2D color plot of the flux from a single spin oriented normally to the ø160 nm SOT loop. **b,** Same as (**a**) but the spin is oriented in the X direction in the plane of the loop. **c,** Same as (**a**) but the SOT diameter is 46 nm. **d, e, f,** Cuts of (**a**), (**b**) and (**c**) at Y = 0. The colored regions above the dotted lines in (**d**), (**e**) and (**f**) show the locations where the Pb SOTs have spin sensitivity better than 1 $\mu_B/Hz^{1/2}$. The arrows show the spatial resolution at full width at half maximum.

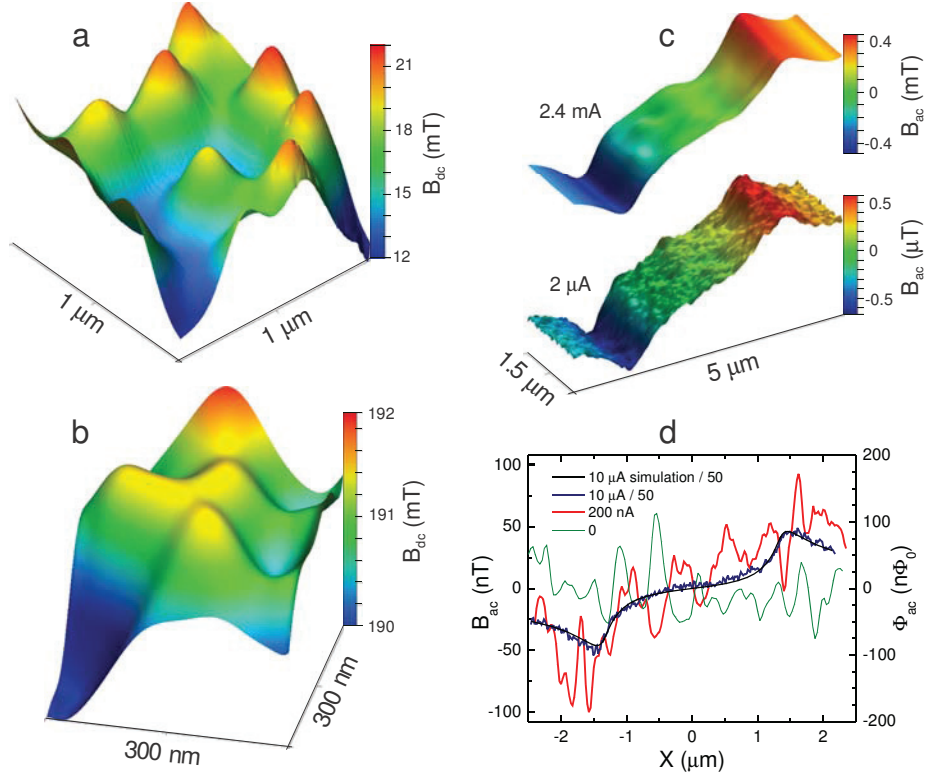

**Figure 5 | Scanning SQUID-on-tip microscopy images of vortex matter (a,b) and of magnetic field distribution generated by AC current (c,d) in Nb film at 4.2 K. a,** 3D false color plot of the magnetic field distribution measured with a ø117 nm Pb SOT displaying pronounced peaks (red) at the vortex centers. The image shows the raw data with no processing measured at a constant height of about 50 nm above the surface on the Nb film ($H$ = 28 mT, scan area 1×1 $\mu m^2$, pixel size 7 nm, and scanning speed 150 nm/sec). The minimum distance between the vortices is 330 nm. Nb film thickness is 420 nm and the penetration depth extracted from the vortex shape is $\lambda \cong 105$ nm. **b,** Similar image as in (**a**) obtained with a ø88 nm Pb SOT in applied field of about 0.2 T. Scan area is 300×300 $nm^2$, pixel size 2 nm, scan speed 40 nm/sec. The distance between the closest vortices is 120 nm. **c,** 3D false color image of $B_{ac}$ measured at ~100 nm above a 3 $\mu m$ wide Nb strip with $I_{ac}$ = 2.4 mA (top) and 2 $\mu A$ (bottom) at 609 Hz using a lock-in amplifier and ø70 nm Pb SOT. Both images show raw unprocessed data (scan area 1.5×5 $\mu m^2$, pixel size 33.5 nm, scanning speed 560 nm/sec, lock-in time constant 100 ms). The wavy features in the top image are due to small local AC displacement of vortices by the $I_{ac}$ = 2.4 mA. **d,** $B_{ac}$ line profiles at $I_{ac}$ = 10 $\mu A$ (blue, divided by 50), 200 nA (red), and zero current (green) (pixel size 25 nm, scanning speed 30 nm/sec, lock-in time constant 1 s). The black curve shows the calculated theoretical field profile (divided by 50) induced by 10 $\mu A$ current at 100 nm above the surface of 400 nm thick superconducting strip. Note some six orders of magnitude variation in the field values between the different panels.

For testing the SOT sensitivity under realistic scanning conditions we imaged the AC magnetic field $B_{ac}$ generated by an AC transport current flowing in a 3 μm wide Nb strip using a lock-in amplifier and a ø70 nm Pb SOT. Figure 5c shows raw data of $B_{ac}$ distributions at AC currents of 2.4 mA and 2 μA. Figure 5d shows some additional line scans: At zero current (green curve) the noise level is fully consistent with our measurement of the white noise power density of ~90 n$\Phi_0$/Hz$^{1/2}$ in this device. The blue data show $B_{ac}$ at 10 μA (divided by 50) along with a theoretically calculated $B_{ac}$ at 100 nm above a superconducting strip carrying 10 μA current showing an excellent fit to the data. The red line shows the signal attained at 200 nA demonstrating the detectable level of AC signal of about 50 nT or 93 n$\Phi_0$ with lock-in time constant of 1 s. As shown in Fig. 4 this level should allow detection of a single electron spin.

The presented results show that the SQUID-on-tip technology has the sensitivity and the spatial resolution that are required for detection and imaging of magnetic moments of individual electrons. With the anticipated further enhancements this method should establish a powerful new tool for quantitative nanoscale scanning magnetic microscopy with single spin sensitivity.

## Methods

Pipettes with tip diameters in the range of 40 to 300 nm were fabricated from quartz tubes with 1 mm outside diameter and 0.5 mm inside diameter using a Sutter Instrument P2000 micropipette puller. Before the deposition of a superconducting film, two gold contacts 200 nm thick were deposited through a mask along the cylindrical part of the pipette, opposite each other. The mask with a 10 mm long and 0.3 mm wide opening prevents an overlap between the two gold electrodes and also prevents gold deposition onto the tapered part of the pipette. The superconducting films were deposited in three steps using in-situ rotators. In the first step the pipette is pointed towards the source (defined as 0° orientation) and a thin film is deposited onto the apex ring of the pipette forming the superconducting loop of the SQUID. The pipette is then rotated to a 105° orientation and an electrode is deposited on one side of the pipette, connecting the apex ring and the gold contact. The third deposition is performed at a -105° orientation, forming the second electrode on the opposite side of the pipette. The order of the three depositions can be changed. The deposition at angles larger than 90° assures formation of a gap that prevents shorting the two leads (see Fig. 1). In the regions where the loop overlaps the leads stronger superconductivity is attained, while the short parts of the loop next to the gaps between the leads act as two Dayem-bridge weak links, thus forming a self-aligned SQUID-on-tip[12,26]. Because of their small size, the SOTs are extremely sensitive to electrostatic discharge. The devices were therefore shorted at all times and opened only during the measurements.

**Nb SOT fabrication:** The conventional method of magnetron sputtering of Nb cannot be applied to our process since the self-aligned fabrication requires deposition from a point source. We have used an alternative approach of e-gun deposition which requires, however, strict ultra-high vacuum (UHV) conditions because of the high sensitivity of Nb to contamination. We have therefore constructed a dedicated UHV system with base pressure of 3.5×10$^{-11}$ Torr and pressure during the e-gun deposition of Nb of 2×10$^{-9}$ Torr. The deposition rate was ~7 Å/s. Even though the $T_c$ of the films was around 9 K, the tips showed finite residual resistance of about 100 Ω at 4.2 K, possibly caused by contamination from the quartz and stress in Nb film. This hurdle was overcome by depositing a 10 nm thick AlO$_x$ buffer layer on the quartz tips, resulting in zero residual resistance and functional SOTs. The thickness of Nb electrodes on the sides of the pipette was typically 35 nm and the thickness of the apex ring was about 23 nm.

**Pb SOT fabrication:** The high surface mobility of the Pb atoms results in island growth and in a high percolation threshold in Pb films deposited onto room temperature substrates. Coupling the tips to liquid nitrogen reservoir improved the film morphology, but still did not allow reliable fabrication of SOTs. We therefore developed a thermal evaporation system in which the pipettes are mounted on the end of a homemade continuous flow $^4$He cryostat that can be rotated around a horizontal axis while in vacuum. The cold end of the cryostat is itself enclosed in a liquid nitrogen jacket. Because of the tip geometry, it is impossible to provide good thermal contact along the whole length of the quartz pipette, so the tip was fitted inside a narrow gap in a brass block mounted onto the cryostat. The system base pressure was 2×10$^{-7}$ Torr and the cryostat temperature was 8 K. Before the deposition, $^4$He gas at 3×10$^{-5}$ Torr was introduced into the chamber to facilitate tip cooling and then pumped out to the base pressure. The deposition rate was ~5 Å/s and film thickness of the Pb SOT was about 40 nm for the electrodes and 25 nm for the apex ring.

**Electrical circuit:** A simplified measurement circuit is shown in Fig. 2a. An input voltage $V_{bias}$ in series with a cold $R_{bias} \cong 5$ kΩ resistor applies current to the SOT that is shunted by a cold $R_{shunt} \cong 3$ Ω. A series SQUID array amplifier[32] (SSAA) is used as a cryogenic current-to-voltage converter to read out the SOT current $I_{SOT}$. The SSAA is inductively coupled to the SOT circuit through $L_{IN}$ and is operated in a closed feedback loop, the output of which $V_{FB}$ is proportional to $I_{SOT}$. When the SOT is in the superconducting state, most of the current supplied by $V_{bias}$ passes through the SOT giving rise to the linear part of the I-Vs in Figs. 2a and 2b. In the resistive state above $I_c$ most of the current flows through $R_{shunt}$ resulting in the negative differential resistance. Nevertheless, the I-Vs may remain nonhysteretic due to the low resistance shunt $R_{shunt}$ (see Supplementary Information).

**Imaging:** An in-house-built scanning stage operating at 4.2 K was used for vortex imaging. The coarse motion was attained by three attocube piezoelectric positioners based on the slip-stick mechanism. The scanning was performed by an attocube integrated xyz scanner with xy range of 30 μm and z range of 15 μm at 4.2 K. The microscope provides for a z-axis feedback based on quartz tuning fork, but for the results presented in this work the SOT was scanned at a constant height above the Nb film with no feedback. The Nb film was sputtered on Si substrate and had a thickness of 420 nm.

## Acknowledgements

This work was supported by the European Research Council (ERC advanced grant) and by the Minerva Foundation with funding from the Federal German Ministry for Education and Research. Y.A. acknowledges support by the Azrieli Foundation and by the Fonds Québécois de la Recherche sur la Nature et les Technologies. M.H acknowledges support from the Weston Visiting Professorship program. E.Z. acknowledges support by the US-Israel Binational Science Foundation (BSF).

## Author contributions

D.V. and Y.A. fabricated and measured the SOT devices. Y.A, M.R., Y.M. and L.N. developed the Pb deposition setup. Y.M, M.R., and D.V. designed and built the Nb evaporator. L.E. designed and constructed the scanning SOT microscope. L.E., D.H and J.C. carried out the magnetic imaging. A.F. and Y.S. contributed to the development of the SOTs and the microscope. M.H. developed the SQUID array readout system. D.V. and E.Z. co-wrote the paper. All authors contributed to the manuscript.

## Additional information

The authors declare no competing financial interests. Correspondence and requests for materials should be addressed to D.V. or E.Z.

# Supplementary Material
# Scanning nano-SQUID with single electron spin sensitivity

**Quantum limit of sensitivity for nano-SQUIDs**

In the case of an optimized SQUID ($\beta_L \approx \beta_C \approx 1$), the quantum noise limit (QNL) is $S_\Phi^{1/2} = (\hbar L)^{1/2}$, where $L$ is the total inductance of the SQUID loop and $\hbar$ is the reduced Plank's constant. A number of SQUIDs have been shown to reach flux noise level close to the quantum limit and each SQUID design would have its own QNL [7, 8]. In general, inductance consists of two parts – geometric and kinetic: $L = L_g + L_k$. In SOTs $L$ is governed by the kinetic inductance $L_k$ [12] while the geometric inductance $L_g$ is very small (for ø200 nm loop $L_g \cong 0.25$ pH and for ø50 nm loop $L_g \cong 0.01$ pH). The total inductance can be extracted from the modulation depth of the critical current with magnetic field that is determined by the parameter $\beta_L = LI_c/\Phi_0$ [7,19]. For the ø160 nm Pb SOT we obtain $L \approx 10.5$ pH while for the ø56 nm Pb SOT we estimate $L \approx 5.7$ pH resulting in QNL of $S_\Phi^{1/2}$ = 16 n$\Phi_0$/Hz$^{1/2}$ and 12 n$\Phi_0$/Hz$^{1/2}$ respectively. The measured spectral flux noise density of 50 n$\Phi_0$/Hz$^{1/2}$ in these devices is thus 3.2 to 4.2 times larger than the QNL. In the Nb SOT the critical current and the modulation depth are much smaller resulting in a significantly higher $L \approx 150$ pH and correspondingly higher QNL of $S_\Phi^{1/2}$ = 60 n$\Phi_0$/Hz$^{1/2}$. This result partially explains the much higher flux noise of 3.6 µ$\Phi_0$/Hz$^{1/2}$ measured in this device. The shot noise limit is given by $S_\Phi^{1/2} = (hL)^{1/2}$ and is $(2\pi)^{1/2}$ times larger than the QNL resulting in $S_\Phi^{1/2}$ = 40 n$\Phi_0$/Hz$^{1/2}$ in the case of ø160 nm Pb SOT and 30 n$\Phi_0$/Hz$^{1/2}$ in ø56 nm Pb SOT. Finally, the thermal or Johnson noise can be calculated [7,8] using the expression $S_\Phi^{1/2} = (4k_B TL(\pi LC)^{1/2})^{1/2}$, where $k_B$ is Boltzmann's constant, $T$ is the temperature, and $C$ is the capacitance of the SOT which we estimate to be 2 fF. Because of the very small inductance and capacitance the thermal noise at 4.2 K turns out to be lower than the quantum noise, resulting in $S_\Phi^{1/2} \approx 12.5$ n$\Phi_0$/Hz$^{1/2}$ and 8 n$\Phi_0$/Hz$^{1/2}$ for the ø160 nm and ø56nm Pb SOTs respectively.

**Measuring the SOT current-voltage characteristics with a series SQUID array amplifier (SSAA)**

The Dayem-bridge-type weak links often have hysteretic current-voltage characteristics (I-Vs) if they are biased with a current source and their voltage is measured. This hysteresis is very inconvenient for sensor applications. We have therefore implemented a quasi-voltage bias scheme, shown in Fig. 2a, in which a small bias resistor (1 to 3 Ω) provides an effective voltage bias while the SOT current is measured using an inductively coupled transimpedance SSAA [4,12,26,32]. In this case, even though the I-Vs have a negative differential resistance region, they can nonetheless be stable (i.e. no switching and no hysteresis) as shown in Figs. 2a and 2b of main text. The advantage of this scheme is demonstrated in Fig. S1 where the same Nb SOT was measured using the two schemes. The usual current bias measurement results in switching and hysteretic I-V characteristics as shown in Fig. S1a while the quasi-voltage bias conditions result in a stable operation and smooth non-hysteretic I-V shown in Fig. S1b. The reason for this difference is the negative differential resistance of the intrinsic I-V characteristics of the SOT. This region of negative differential resistance is unstable if current is imposed, but can be stabilized by imposing a voltage.

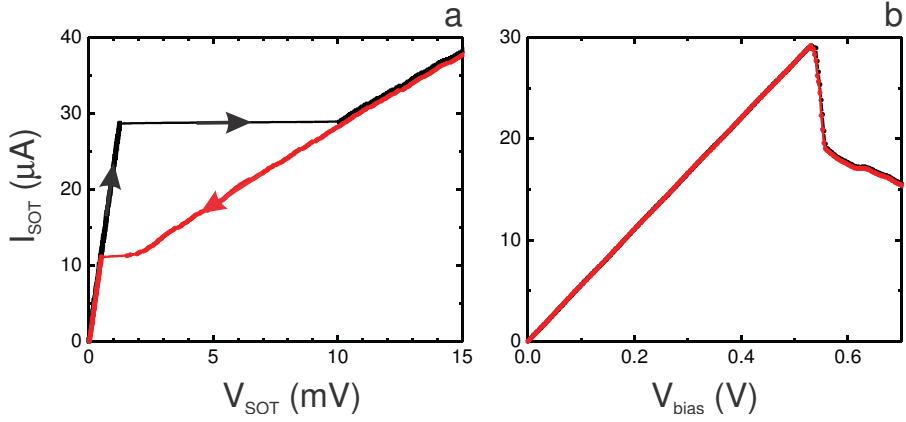

**Figure S1 | Examples of I-V characteristics of Nb SOT measured using different types of bias conditions at 4.2K. a**, The usual current bias condition results in switching and hysteretic behavior. Note that the jump to the normal branch results in $V_{SOT}$ that is larger than $2\Delta/e$ due to the heating of the tip in the normal state. **b**, Quasi-voltage bias using the circuit of Fig. 2a with series SQUID array amplifier results in reversible non-switching characteristics. Black and red colors denote voltage (or current) ramping up and down correspondingly.

**Noise measurements**

To further demonstrate the direct relation between the measured flux noise spectral density and the ability of the device to measure real signal we performed an additional experiment in which the SOT was placed inside a small superconducting coil excited by an AC transport current. First we measured the noise spectral density of a Pb SOT with a spectrum analyzer without exciting the coil showing a white noise level of below 80 $n\Phi_0/Hz^{1/2}$ (Fig. S2a). After that we activated the coil and ran a lock-in measurement with a time constant of 1 s and excitation frequency of 660 Hz varying the amplitude of the AC field to find out the lowest detectable signal of the SOT as shown in Fig. S2b. It is clear from the figure that the minimum detectable signal of the SOT is consistent with the noise measurements of the spectrum analyzer.

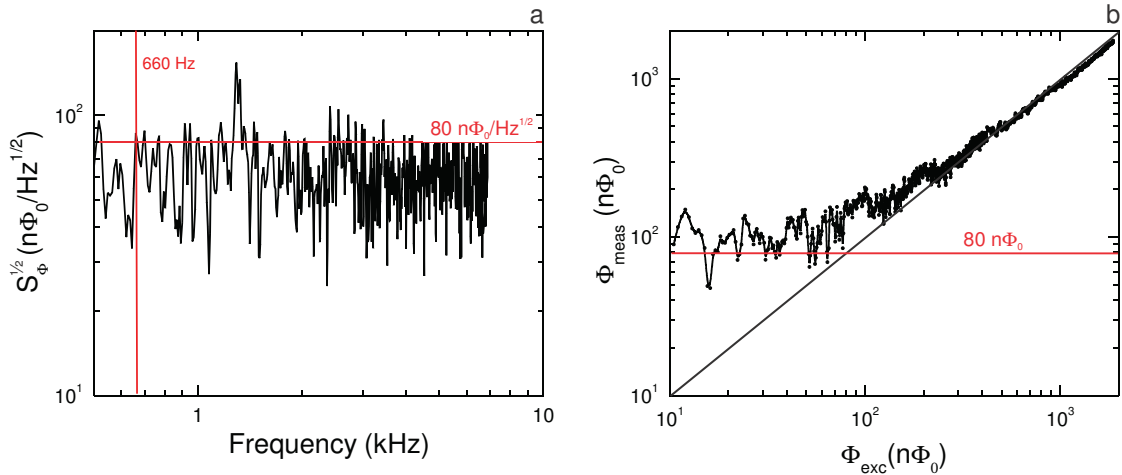

**Figure S2 | Comparison of the noise measurement in a Pb SOT using two techniques at 4.2K. a**, Spectrum analyzer measurement showing white flux noise level of below 80 $n\Phi_0/Hz^{1/2}$. **b**, Lock-in amplifier measurement of the AC flux in the SOT vs. the excitation flux generated by a small coil. The time constant of the lock-in was set to 1s and excitation frequency to 660 Hz. The red lines are guide to the eye showing noise level of 80 $n\Phi_0$.

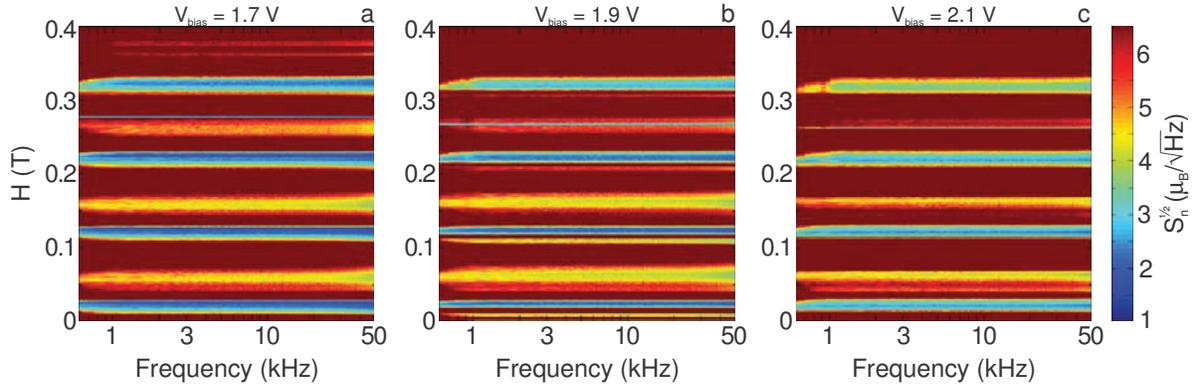

**Figure S3 | Spin noise spectra of ⌀160 nm Pb SOT vs. applied field at 4.2K measured at different constant biases. a**, $V_{bias}$ = 1.7 V. **b**, $V_{bias}$ = 1.9 V. **c**, $V_{bias}$ = 2.1 V.

**Influence of the SOT bias on the spin sensitivity**

Figure S3 complements the data of Fig. 3c showing the spin noise spectra vs. applied field at various constant biases demonstrating the ability of the SOT to operate over a wide range of biases and magnetic fields with comparable performance.

**Influence of the SQUID asymmetry on the working range of the SOT**

Smaller SQUID sizes should mean generally higher operating fields. However the central lobe of the interference pattern can be shifted from zero field if the critical currents or the inductances of the two junctions are not equal [7, 19]. Figure 2c of main text demonstrates a highly asymmetric interference pattern of ⌀238 nm Nb SOT with a sensitive region around zero field. The asymmetry of the interference pattern can be controlled intentionally by adding an extra angle during the deposition as shown schematically in Fig. S4a, resulting in unequal lengths of the two bridges and correspondingly unequal critical currents and inductances. Figure S4b shows critical current interference patterns of an asymmetric ⌀119 nm Pb SOT at positive and negative bias polarities. The SOT asymmetry causes the two patterns to be shifted in field in opposite directions with respect to zero field. Figure S4c shows the magnetic field response of the SOT at constant positive and negative biases (dashed) with the larger of the two responses emphasized by the solid line. This figure demonstrates that by choosing the appropriate polarity an asymmetric SOT can operate continuously over a very wide range of fields. In this device the lowest response of about 1.2 V/T is obtained at zero field. The corresponding noise measurements show that even at this worst working field the spin sensitivity of this device was 25 $\mu_B$ /Hz$^{1/2}$, which is better than in any previously reported nanoSQUIDs.

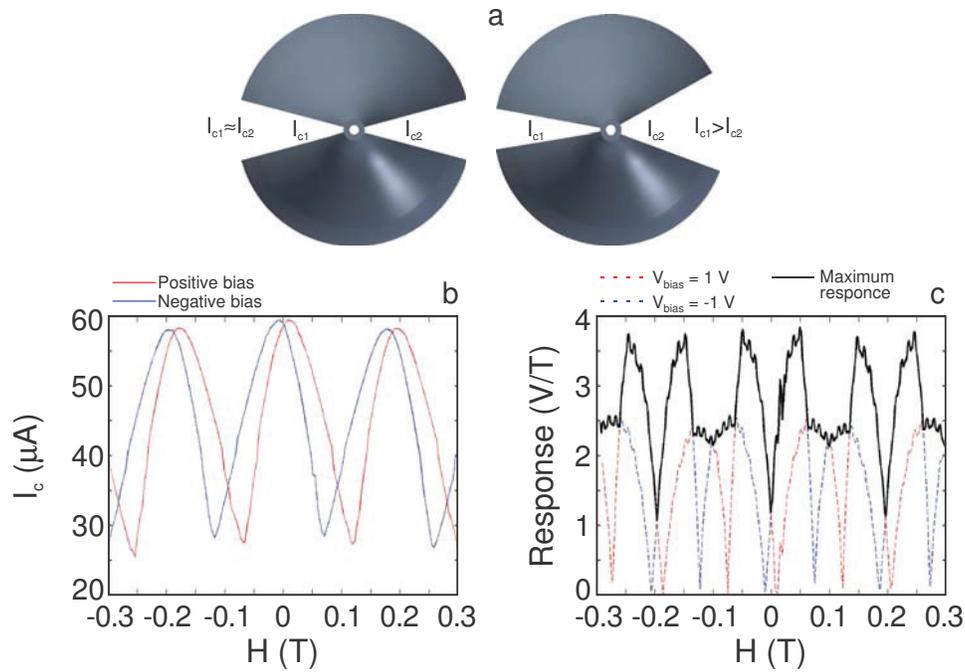

**Figure S4 | Influence of asymmetry on the SQUID interference pattern. a,** A schematic head-on view of the SOT showing a controlled way of introducing the asymmetry by adding a pitch angle during the deposition. **b,** An example of an asymmetric $I_c$ interference pattern in a ø119 nm Pb SOT at opposite bias polarities, demonstrating an opposite shift of the interference patterns with respect to zero field. **c,** Magnetic field response as a function of applied field at positive and negative biases (dashed). The solid line is the combined maximum of the two curves showing a finite response at all fields.